\documentclass{emulateapj}
\usepackage{natbib}
\usepackage{amsmath}
\usepackage{float}
\usepackage{subfigure}
\usepackage{xcolor}
\usepackage{mathrsfs}
\usepackage{hyperref}

\def\medd{\dot M_{\rm Edd}}
\def\ledd{L_{\rm Edd}}
\def\mdot{{\dot M}}
\def\msun{M_{\odot}}
\def\msunyr{M_{\odot}~\rm yr^{-1}}
\def\ergs{\rm ergs~s^{-1}}
\def\mhcm{\rm m_{H}~cm^{-3}}
\def\cms{\rm cm^{-2}}
\def\cmc{\rm cm^{-3}}
\def\dyncm{\rm dyne~cm^{-2}}
\def\kms{\rm km~s^{-1}}
\def\be{\begin{equation}}
\def\ee{\end{equation}}
\catcode`\@=11  

\shorttitle{NLR Outflows in NGC 4151}
\shortauthors{Mou et al} 

\begin{document}
\bibliographystyle{apj}
\title
{Numerical Study on Outflows in Seyfert Galaxies I: Narrow Line Region Outflows in NGC 4151}

\author {Guobin Mou \altaffilmark{1,2}, Tinggui Wang\altaffilmark{1,2}, Chenwei Yang\altaffilmark{1,2}.}
 \altaffiltext{1}{CAS Key Laboratory for Research in Galaxies and Cosmology, Department of Astronomy, University of Science and Technology of China, Hefei 230026, China; gbmou@ustc.edu.cn}
 \altaffiltext{2}{School of Astronomy and Space Science, University of Science and Technology of China, Hefei 230026, China}

\begin{abstract} 
The origin of narrow line region (NLR) outflows remains unknown. In this paper, we explore the scenario in which these outflows are circumnuclear clouds  driven by energetic accretion disk winds. We choose the well-studied nearby Seyfert galaxy NGC 4151 as an example. By performing 3D hydrodynamical simulations, we are able to reproduce the radial distributions of velocity, mass outflow rate and kinetic luminosity of NLR outflows in the inner 100 pc deduced from spatial resolved spectroscopic observations. The demanded kinetic luminosity of disk winds is about two orders of magnitude higher than that inferred from the NLR outflows, but is close to the ultrafast outflows (UFO) detected in X-ray spectrum and a few times lower than the bolometric luminosity of the Seyfert. Our simulations imply that the scenario is viable for NGC 4151. The existence of the underlying disk winds can be confirmed by their impacts on higher density ISM, e.g., shock excitation signs, and the pressure in NLR. 
\end{abstract} 


\section{INTRODUCTION}
AGN outflows may play important roles in regulating the growth of supermassive black holes (SMBHs) and evolution of their host galaxies (\citealt{Silk1998}; \citealt{DiMatteo2005}; \citealt{Hopkins2010}; \citealt{Gaspari2011}; \citealt{Fabian2012}; \citealt{King2015}).  
These outflows are found to be ubiquitous, widely opened, and sometimes very energetic, suggesting that potentially, they can efficiently eliminate and heat interstellar medium (ISM) or intracluster medium. 
Manifesting as blueshifted absorption/emission lines, a variety of AGN outflows have been detected in various bands, and span over many magnitudes of spatial scales from a few tens of Schwarzschild radius to $10^{4}$ pc.
With \emph{XMM-Newton} surveys, \citet{Tombesi2010} reported that about one-third of radio-quiet AGNs show highly ionized absorbers with outflow velocities higher than $10^{4}~\kms$ (ultra-fast outflows, UFOs), which are probably located at $10^{2-4}~R_s$, and majority of UFOs may contribute significantly to feedback (\citealt{Tombesi2013}).  These results are further confirmed by later \emph{Suzaku} detections (\citealt{Gofford2013, Gofford2015}). \citet{Gibson2009} reported BAL QSOs with blueshifts of $10^{3-4} ~\kms$ in optical/UV band, of which the outflows may be located at $10^{2-3}$ pc (He et al. in preparation). \citet{Cicone2014} reported massive molecular outflows with blueshifts of $10^{2-3}~\kms$ by CO millimeter emission, of which the outflow extensions are of kiloparsec scale. 
While AGN outflows detected in the accretion disk scale represent disk winds themselves, those in galactic scale may be performance of feedback, including ISM accelerated by AGN radiation or dragged by disk winds (e.g., \citealt{Thompson2015}; \citealt{Costa2014}; \citealt{Wagner2013}; \citealt{ZubovasKing2013}). 

Supplying a way to study the interaction between accretion disk and the ambient ISM, the intermediate scale (between accretion disk scale and galactic scale) of NLR provides an indirect solution to explore the physics of unresolvable accretion process, and a direct solution to study AGN feedback in host galaxies. High resolution observations have revealed outflow features of NLR in many Seyfert galaxies, and have made constraints on the kinematics of NLR outflows (e.g., \citealt{Crenshaw2003}; \citealt{Barbosa2009}).

Among bright AGNs, NGC 4151 is an outstanding object, since it is the closest Seyfert 1 galaxy to us, with a distance of 13.3 Mpc (\citealt{Mundell2003}, 1'' corresponding to 64 pc). The bolometric luminosity of NGC 4151 is 7$\times 10^{43}~\ergs$ according to the measured $\lambda F_{\lambda}(5100\AA)$ (\citealt{Kaspi2005}). The mass of the central black hole is $4\times10^{7} M_{\odot}$(\citealt{Bentz2006}). Therefore the present Eddington ratio of this source is $\sim 1\% ~\ledd$. 
Multi-waveband observations have been carried out during the past tens of years. We here briefly summarize the physics of observed winds/outflows in NGC 4151 from accretion disk scale to NLR scale. 

In the inner part of accretion disk, UFO has been detected with Fe K absorption line, which shows a blueshift of 0.1c (\citealt{Tombesi2010,Tombesi2011}). The ionization parameter is $\lg \xi \sim 4.4$ ($\xi \equiv L_{ion}/n_{H}r^{2}$, $L_{\rm ion}$ is the ionization luminosity integrated from 13.6 eV to 13.6 keV), and column density is $N_{H} \sim 8 \times 10^{22} ~\cms$. UFO may be located at a distance of $4\times10^{14}-6\times10^{15}$ cm (30-500 Schwarzschild radius) from the center (\citealt{Tombesi2012}). By these parameters, the mass outflow rate of UFO is estimated to be in the range of $3\times 10^{-3}-8\times 10^{-2} \msunyr$, and the kinetic power of UFO is $10^{42-43}~\ergs$ (\citealt{Tombesi2010,Tombesi2012}).

In the outer region of accretion disk or in the vicinity of putative torus, X-ray and UV absorption lines reveal another kind of outflows, with a lower blueshifts of $\la 1000~\kms$ (\citealt{Kraemer2005,Kraemer2006}; \citealt{Crenshaw2007}). These absorbers are a blend of several components with different column densities and ionization parameters. 
Locations of these absorbers are still unclear, and based on the column densities and ionization parameters (\citealt{Kraemer2006}), an upper limit of the distance can be obtained: 0.3 pc for highly ionized X-ray absorbers, and 5 pc for the strongest UV absorber -- D+Ea component. 
\citet{Couto2016} suggested that, the distance of highly ionized X-ray absorber may be $\sim 0.01-0.03$ pc, while it is $\sim 0.1$ pc for D+Ea component. 
\citet{Kraemer2005} and \citet{Couto2016} argued that, while it is possible for D+Ea absorber in UV band to be driven by radiation, X-ray absorber is not likely accelerated by radiation, since the relatively high ionization parameter results in a low line force multiplier, further leading to an insufficient radiation driven force with such a low luminosity. The mass outflow rate and kinetic luminosity of these absorbers are estimated to be $0.19 \msunyr$ and $\sim 10^{40}~\ergs$ (\citealt{Crenshaw2007}).

Further out in the NLR at a distance of tens to hundreds parsecs from the center, two extended structures stretching along the SW and NE direction can be seen by [O III] emissions. These structures are also spatially overlapped with near-infrared structures and X-ray structures (\citealt{StorchiBergmann2009, StorchiBergmann2010}; \citealt{WangJF2011b, WangJF2011c}). According to \citet{Crenshaw2010}, the geometry of NLR is a cone with half opening angle of $33^{\circ}$, and the inclination angle is $45^{\circ}$ ($0^{\circ}$ corresponds to the cone-axis in the plane of the sky). \citet{Kraemer2000} successfully generated a photoionization model to estimate the serial NLR emission lines in optical band, in which they conclude that the emission line gas is inhomogeneous, and consists of a dense component and a tenuous component. Later, by analyzing the high resolution spectra in optical and near infrared bands, \citet{Das2005} found that in some locations there are two or more components with different velocity centroids, which means one projected position actually contains several NLR clouds with different kinematic characteristics (see also \citealt{StorchiBergmann2010}). 
Besides, by analyzing infrared data, \citet{StorchiBergmann2009} find evidences for shock excitation effect, i.e., [Fe II]/Pa$\beta$ is significantly enhanced beyond the inner 0.6'' on both sides of NLR, indicating that shock excitation plays a important role there. The velocity of NLR outflow increases linearly from 0 at $r\sim 0$ pc to 800 $\kms$ at distance of $r=96$ pc, and then turns over to decrease linearly to 0 at $r=288$ pc (\citealt{Das2005}), resulting in a similar behavior of mass outflow rate and kinetic luminosity. The mass outflow rate increases with distance to a maximum of $\sim 3~\msunyr$ at $r \sim 70$ pc, and then turns over to decrease with distance, while the kinetic power shows a maximum of $\sim 4\times 10^{41}~\ergs$ at $r \sim 90$ pc (\citealt{Crenshaw2015}). Such a kind of kinematic feature is also detected in some other nearby Seyferts, such as NGC 1068 (\citealt{Crenshaw2000}), Mrk 3 (\citealt{Ruiz2001}), and Mrk 573 (\citealt{Fischer2010}).

\citet{Everett2007} proposed an analytic Park wind model to explain kinematics of NLR outflows, in which Park wind is generated by thermal expansion induced by heating of central continuum. NLR outflows are corresponding to the thermal winds themselves or clouds dragged by the thermal winds, while deceleration of NLR outflows is due to drag forces from gravity and resistance of ISM when NLR outflows passing it. Unfortunately, the strong adiabatic cooling of Park wind makes it not successful to match the overall velocity profiles. 
\citet{Das2007} proposed another analytic model in which the NLR clouds are directly accelerated by AGN radiation pressure. However, the velocities reach to the maximum too sharply to match the observed kinematics. 
We also note that some numerical simulations which may show hints on the origin of NLR outflows. \citet{Moscibrodzka2013} and \citet{Barai2012} studied the accretion flow properties in NLR scale by considering the X-ray heating from the central source and cooling effects. They found that initially smooth accretion flow gradually develops into thermal instabilities and convection instabilities. Some of the formed clouds move outwards resembling the NLR outflows. However, velocities of outwards clouds are actually too low to account for observations.

Abundant observations of NGC 4151 make it possible to explore the formation of outflows in different scales and the link between them.
Taking into account that the velocity and mass outflow rate increase with distance in the inner $\sim 100$ pc (\citealt{Hutchings1999}; \citealt{Das2005}), NLR outflows do not seem to originate from the accretion disk, but must be generated by circumnuclear ISM (mainly clouds) driven by disk winds or by radiation. In this paper, we explore the first scenario -- clouds driven by disk winds, and leave the second scenario to the future work.  

We introduce disk winds in section 2, and show simulation details in section 3. We present the simulation results in section 4, and make discussions and summarize our results in section 5 and section 6.


\section{Disk Winds} \label{diskwinds}
Accretion disk around a black hole has been studied during the past tens of years. So far, it is widely accepted that there are three accretion modes: slim disk for Eddington or super Eddington accretion rate, standard thin disk for moderate accretion rate, and hot accretion flow for low accretion rate (see review by \citealt{YuanNarayan2014}). There is a critical value of the mass accretion rate between thin disk and hot accretion flow, and specifically, transition appears between these two accretion modes when $\mdot_{acc} \sim 10^{-2} ~\medd$ (\citealt{Xie2012}; also see \citealt{Done2007} and \citealt{Zdziarski2004} for the case of stellar mass BH.). 

For NGC 4151 with luminosity of $\sim 1\% \ledd $, the mass accretion rate is just in the critical range. A question arises: what is the accretion mode in this source?
The ``big blue bump'' in intrinsic extreme-UV continuum is usually regarded as a signature of standard thin disk. However, whether this bump exists in NGC 4151 is still unknown yet. \citet{Schulz1993} argued that the ``big blue bump'' may exist for interpreting the extended emission-line region ($\geq$500 pc), while \citet{Alexander1999} argued that the SED which best reproduces the NLR line emission in the 100$\sim$500 pc scale does not show that bump. Nevertheless, the certain existing of UV radiation indicates that the standard thin disk indeed exists.  
Besides, for such a low Eddington ratio, \citet{Yuan2004} concluded that the transition radius may exist, and may be several tens of Schwarzschild radius, inside which the disk is a hot accretion flow. \citet{Lubinski2010} analyzed the X-ray data of NGC 4151, and found that the relatively weak strength of reflection from the disk can be explained by an inner hot accretion flow surrounded by a cold disk truncated at $\sim15~r_g$ ($r_g \equiv GM_{bh}/c^2$). However, \citet{Keck2015} claimed that they do not find evidence for a truncated disk from spectral fitting of their observations.  
In short, we argue that the accretion disk in NGC 4151 should be a thin disk, or ``truncated thin disk+ inner hot accretion flow'' (see Figure 1 in \citealt{Esin1997}). 

For a thin disk, winds can be launched under line-driven force with luminosity $\ga 0.1~\ledd$, as revealed by hydrodynamic simulations (\citealt{Proga2000, Proga2004}; \citealt{Nomura2016}) or non-hydrodynamic studies (\citealt{Risaliti2010}). These studies also found that, for a luminosity as low as $L \sim 0.01~\ledd$, disk winds can not be launched because the line-driven force is too weak to overcome the gravity of BH. However, this may be not the real case. Theoretically, magnetic driven force has not been considered in these studies, which may play an important role in launching winds in the case of low luminosity. On the other hand, observations suggest that winds do exist in AGNs as dim as $\sim 10^{-2} ~\ledd$, such as NGC 4151, NGC 3783, NGC 3516, NGC 5548, and NGC 4051, and the velocities of winds can be as large as 0.1c for NGC 4151, and $\sim 5000 ~\kms$ for the last four sources (the first three refers to \citealt{Tombesi2010}; NGC 5548 refers to \citealt{Kaastra2014}; NGC 4051 refers to \citealt{Peterson2004}; Eddington ratios refer to \citealt{Vasudevan2009}).

For a hot accretion flow, simulations have shown that disk winds can be produced under the buoyant force (\citealt{YuanWB2012a}) or magnetic centrifugal force (\citealt{YuanBW2012b}) in 2-dimensional hydrodynamic (HD) and 2-dimensional magnetohydrodynamic (MHD) case respectively, or common driving forces of magnetic field pressure and thermal pressure in a more recent 3-dimensional General Relativity-MHD study (\citealt{Yuan2015}). Therefore, if an inner hot accretion flow indeed exists in NGC 4151, the disk wind launched from hot accretion flow may account for the UFO component. 
 
Although for NGC 4151, the UFO component in disk winds has been detected, the integrated disk winds that drive clouds in our model are still unclear, since they are a blend of disk winds launched at different radii with different velocities. Before performing our studies, we need to know the approximate range of parameters of the disk winds at the inner boundary of simulation domain ($\sim 1$ pc, outside the accretion disk). We argue that, disk winds may be a mixture of UFO generated in the inner disk of $\sim 10^{2} R_{s}$ and slower components generated at larger radii. We note that the maximum blueshift velocity of UV absorber is 1400 $\kms$ (D+E component, \citealt{Crenshaw2007}), and blueshifts of some NLR clouds can be as high as 1200-1500 $\kms$ (\citealt{Hutchings1999}). These can be regarded as a lower limit for disk wind velocity in our model. 
Referring the velocities detected in other similar sources mentioned above, assuming an overall velocity of several thousands $\kms$ is not unreasonable. 
Therefore, considering the observations of other Seyferts with similar activities and the constraints on the velocity in NGC 4151, we set the overall velocity of disk winds at $\sim 1$ pc to be 8000 $\kms$, and we also test another lower velocity of 4000 $\kms$, as shown in Table 1. We regard the kinetic luminosity of UFO as an lower limit for the disk winds, and set the power of disk winds in the order of $10^{43}~\ergs$.

We use photo-ionization code CLOUDY (version 13.03, \citealt{Ferland2013}) to calculate the wind condition in our model. CLOUDY self-consistently treats the ionization and recombination, temperature, radiative transfer process and line emission in the gas and we check the resultant wind temperature in each condition. Considering the incompleteness of the ionization spectrum in NGC 4151, we choose the SED of anther source as our input ionizing spectrum -- NGC 5548 (\citealt{Mehdipour2015}), of which both the SMBH mass and the bolometric luminosity are almost same as NGC 4151 (\citealt{Denney2010}). We calculate two luminosities $L_{\rm bol}=10^{43.5}$ and $10^{44}~\ergs$, with different wind densities. 
For the disk winds with a density scaling as $r^{-2}$, by assuming that the metallicity is solar abundance, we find that the temperature is  $10^{6-7}$ K ($10^{5}$ K for run C), and does not vary with distance. 
When adiabatic cooling is included, the maximum temperature would be lowered by more than one order of magnitude (\citealt{Chelouche2005}). 
However, since the internal energy of disk winds is much lower than the kinetic energy, the effects of temperature on the results is negligible, which is verified in our simulation tests. We here simply force the disk winds to be $10^{6.0}$ K (except the higher wind density case in run C, $10^{5.0}$ K instead). 
\footnote{We have not included the Compton heating and photoionization heating process in simulations. Therefore, without this handle, the temperature of disk winds would be extremely low when expanding into the scale of NLR in the role of line cooling and adiabatic cooling, which is not real. }

Such a kind of disk winds has not been detected in NGC 4151 yet. We argue that, in parsec or sub-parsec scale, this may be caused by the high ionization parameter, e.g., $\log U \sim 2$ for parameters in run A, where $U \equiv Q_{\rm ion}/4\pi cn_{H}r^{2}$, where $Q_{ion}$ is the ionizing photons per second integrated from 13.6 eV to infinity. Another reason may be that, the absorption lines of disk winds expected to exist is obscured by the putative torus or the dusty inner galactic disk (see the geometric mode in \citealt{Crenshaw2010}).

 \begin{figure*}
   \centering
   \includegraphics[width=0.31\textwidth]{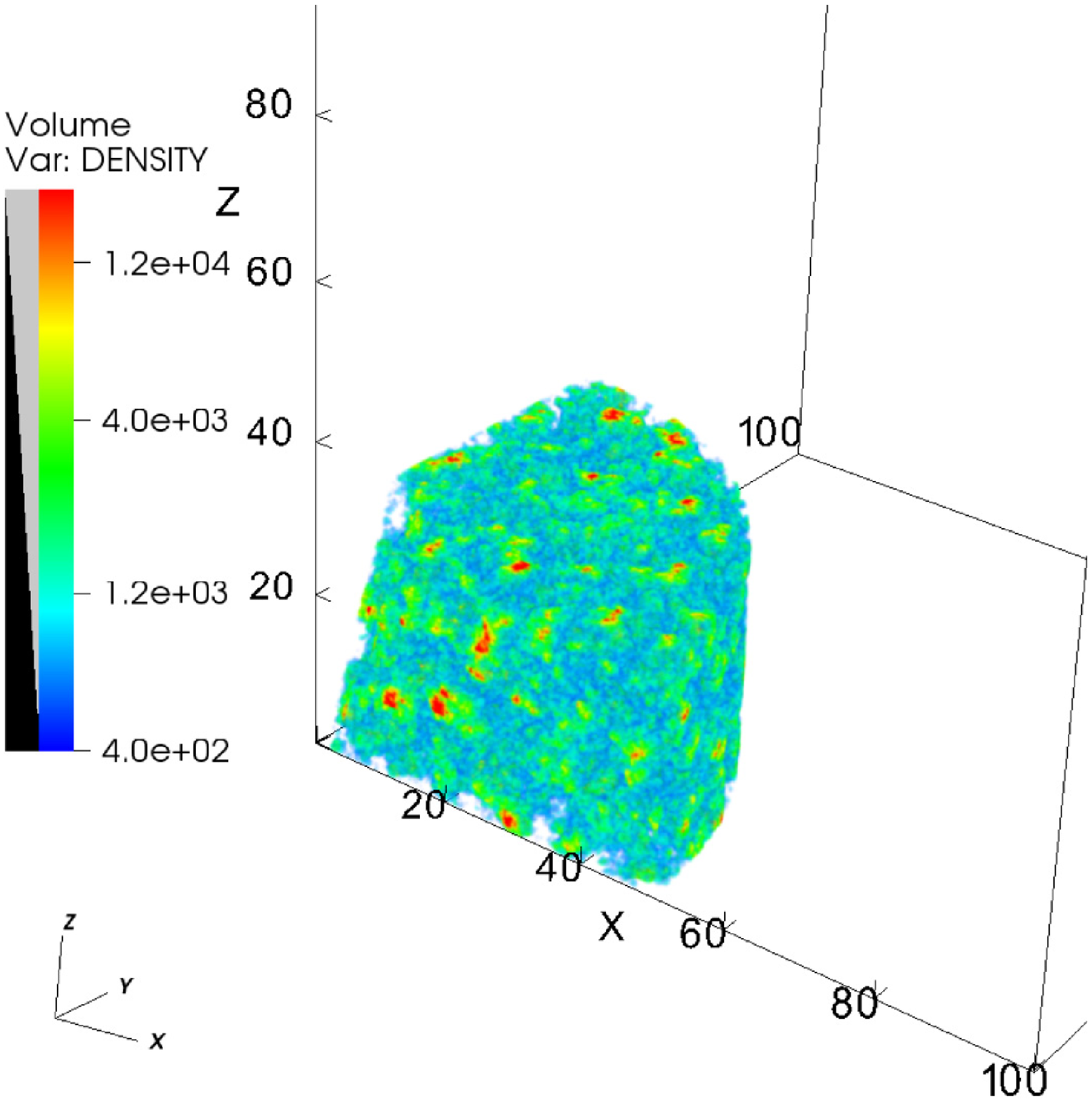} \hspace{0.4cm}
   \includegraphics[width=0.31\textwidth]{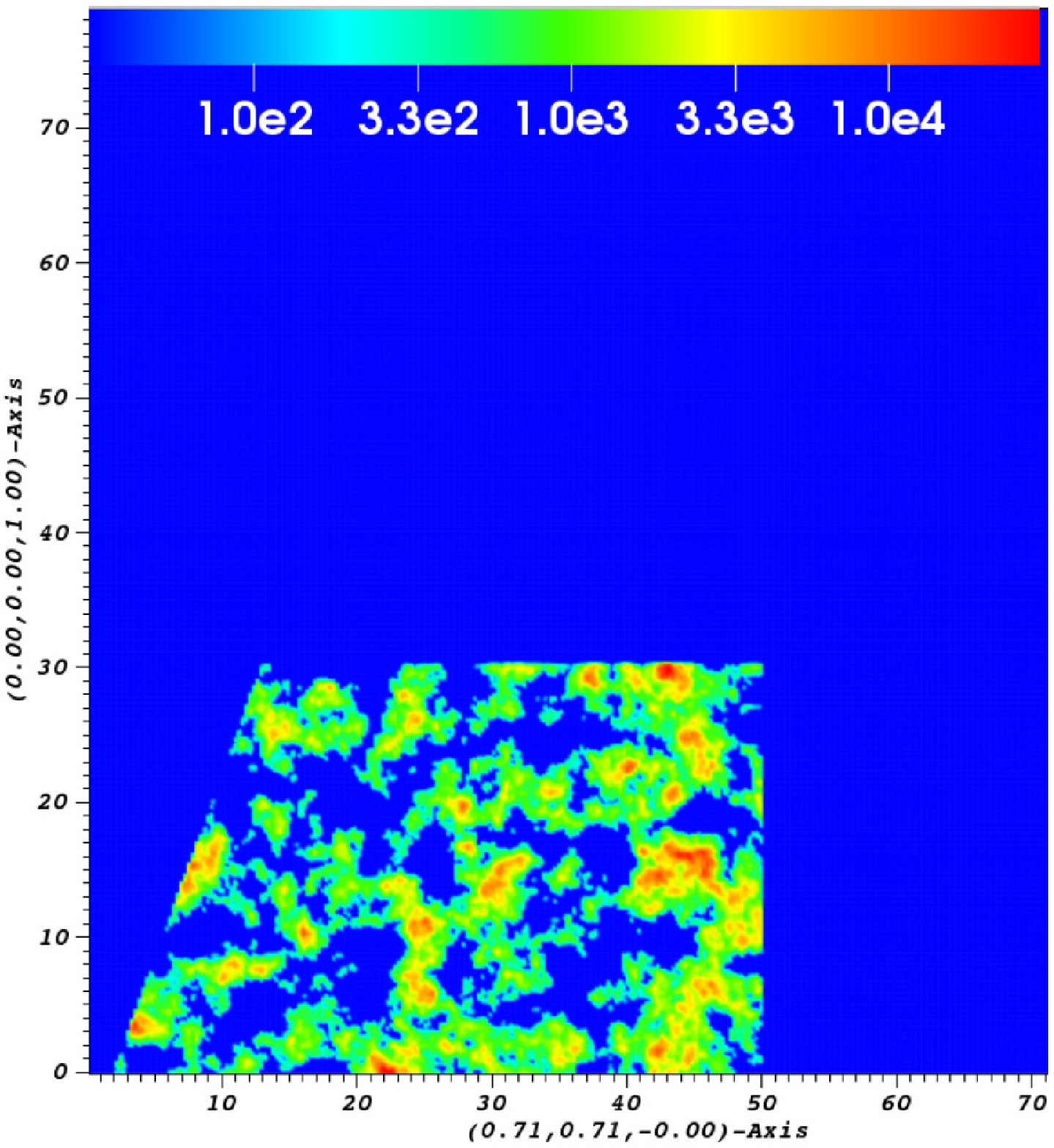}
   \includegraphics[width=0.31\textwidth]{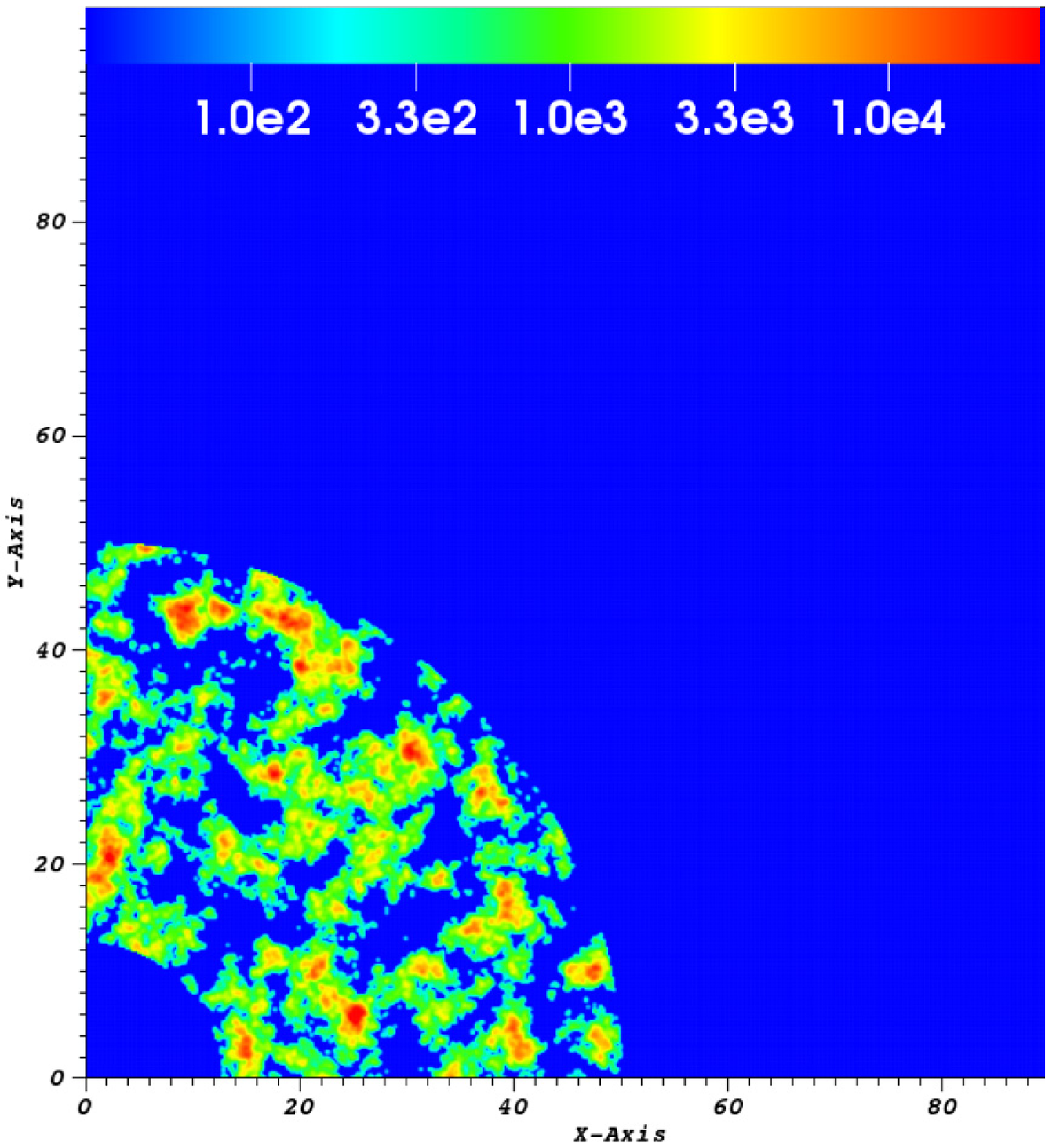}
   \caption{Density field of  circumnuclear clouds in simulation domain (\emph{left}), and in two slices of $\phi=45^{\circ}$ (\emph{middle}), and $z=30$ pc (\emph{right}). When $\phi$ is mentioned in the text, we regard $Z$-axis as the polar axis, and the slice of $Y=0$ as $\phi=0$ by default. Colorbar in each panel represents density in units of $\mhcm$. }
   \label{plot1}
  \end{figure*}

\begin{table*}
  \centering
  \begin{minipage}{130mm}
  \renewcommand{\thefootnote}{\thempfootnote}
  \caption{{\rm  Simulations Parameters}}
  \centering
  \begin{tabular}{@{} c  c  c  c  c  c  c  c  }
  \hline
        & {$\mu$}
        & {$\lambda_{\rm max}$ }
        & {$v_{\rm wind}$}
        & {$\rho_{\rm wind}$ }
        & {$\dot M_{\rm wind}$ }
        & {$L_{\rm kin}$ }
        & {$t$ }
        \\
    Run & (${\rm cm^{-3}}$) &  (pc)  & ($\kms$) & ($\mhcm$) & ($\msunyr$) & ($\ergs$) &  kyr    \\
    (1)   &  (2)                       &  (3)    & (4)           &  (5)            &  (6)               & (7)            & (8)  \\
    \hline
     A & $1.0\times 10^{3}$ & 5      & 8000 & $2\times 10^{3}$   & 1.2 & $2.4\times 10^{43}$  & 140  \\ 
     B & $5.0\times 10^{2}$ & 5      & 8000 & $2\times 10^{3}$   & 1.2 & $2.4\times 10^{43}$  & 120  \\ %
     C & $1.0\times 10^{3}$ & 5      & 4000 & $8\times 10^{3}$   & 1.2 & $1.2\times 10^{43}$  & 170   \\ %
     D & $1.0\times 10^{3}$ & 5      & 4000 & $2\times 10^{3}$   & 0.3 & $0.3\times 10^{43}$  & 250   \\ %
     E & $1.0\times 10^{3}$ & 12.5 & 8000 & $2\times 10^{3}$   & 1.2 & $2.4\times 10^{43}$  & 160   \\ %
    \hline
 \label{table1}
  \end{tabular}
  \tablecomments{(1) model names; (2) mean density of clouds; (3) maximum size of clouds; (4) velocity of injected winds; (5) density of winds at $r=1$ pc; (6) total mass outflow rate of winds; (7) kinetic luminosity of winds; (8) simulation time.} 
 \end{minipage}
\end{table*}


\section{Numerical SIMULATION}

 \subsection{Equations}
The hydrodynamic equations are as follows,
\begin{gather}
  \frac{d \rho}{d t} + \rho \nabla \cdot {\bf v} = 0,\label{hydro1} \\
  \rho \frac{d {\bf v}}{d t} = -\nabla P -\rho \nabla \Phi ,\label{hydro2}  \\
  \frac{\partial e}{\partial t} +\nabla \cdot(e{\bf v})=-P\nabla \cdot {\bf v}+\mathcal{L}_{c}. \label{hydro3}
 \end{gather}
\\
Here $P=(\gamma-1)e$ is thermal pressure, in which $\gamma=5/3$ is the adiabatic index for ideal gas.  
The cooling function $\mathcal{L}_{c}= n_{i}n_{e} \Lambda_{N}$ incorporates free-free and line emission process. We assume the solar abundance for NGC 4151 as in \citet{Kraemer2000, Kraemer2008}, therefore $n_{i}n_{e}=0.652(\rho/m_{p})^{2}$. Following \citet{Sutherland1993}, we fit the collisional ionization equilibrium (CIE) normalized cooling function $\Lambda_N$ as (temperature T in units of keV): 

1. for T $>$ 0.02,   
\be
\begin{split}
\Lambda_{N}=& (6.7\times 10^{-2} T^{-1.2}+7.5 \times 10^{-2} T^{0.5} \\
                       & +2.5\times 10^{-2}) \times 10^{-22} ~{\rm erg~cm^{3}~s^{-1}} ;
\end{split}
\ee

2. for 0.006  $\leq$ T $\leq$ 0.02, 
\be
\Lambda_{N}=1.0 \times 10^{-21} ~{\rm erg~cm^{3}~s^{-1}} ;
\ee

3. for  0.001  $\leq$ T $<$ 0.006,
\be
\Lambda_{N}=2.7\times 10^{-17} T^{2.0} ~{\rm erg~cm^{3}~s^{-1}} .
\ee

We set a ground temperature of 0.001 keV ($\sim 1 \times 10^{4}$ K, below which the cooling term in the code is closed.

We set the gravitational force $\nabla \Phi $ to be zero, as in some simulations with two-phase ISM (\citealt{Wagner2012}, \citealt{Bieri2016}), and this simplification is reasonable since the free fall timescale of clouds is much longer than the dynamical timescale. 

 \subsection{Simulation Setup}

We adopt ZEUSMP code to solve the above equations (\citealt{Stone92}; \citealt{Hayes2006}), and choose 3-D Cartesian coordinates. In order to maintain a high resolution while taking into account the super-computer's memory size, we simulate one octant here (we choose the (+,+,+) octant, while there are 8 octants in total for 3-D Cartesian coordinates). When we analyze the total mass outflow rate and kinetic power, we assume that the other 7 octants are same as the one we simulate. The simulation range is: $0\sim +100$ pc in $X$-direction, $0 \sim +100$ pc in $Y$-directions, and $0 \sim +140$ pc in $Z$-direction, and $Z$-axis stretches along the axis of NLR. The SMBH is just located at the origin. The boundaries of $X=0$, $Y=0$, $Z=0$ are set to be inflecting boundary condition, while the boundaries of $X=100$ pc, $Y=100$ pc and $Z=140$ pc are set to be outflow boundary condition. The computation domain is uniformly divided into 500, 500 and 700 meshes in $X$-, $Y$-, $Z$-direction, respectively, and each mesh is a cube with a side length of 0.2 pc. It takes 3000 CPU-hours for the basic run. 
Besides, when the term of $r$, $\theta$, or $\phi$ is mentioned in the text, we regard $Z$-axis as the polar axis, and the slice of $Y=0$ as $\phi=0$ by default.
  
 \subsection{Initial Conditions}

The initial gas distribution contains two components: the hot diffuse gas and the cold dense clouds. The density of hot diffuse gas is set to be a constant of 3 $\mhcm$, with a uniform temperature of $1.4\times 10^{6}$ K. The density distribution of the clouds is assumed to be random field that satisfying single-point lognormal statistics and two-point fractal statistics (Lewis \& Austin 2002). 

The lognormal probability function of the density P($\rho$) is 

\be
 P(\rho) =\frac{1}{s \sqrt{2\pi} \rho} {\rm exp} \left[\frac{-({\rm ln} \rho -m)^2}{2 s^2} \right] ,
\ee
in which the mean $\mu$ and the variance $\sigma^2$ of the density are given by: 
\be
m={\rm ln} \frac{\mu^{2}}{\sqrt{\mu^2+\sigma^2}}, ~~ s=\sqrt{\rm ln \left(\frac{\sigma^2}{\mu^2}+1 \right)} .
\ee

We set $\sigma^2=5 \mu^2$, which is consistent with the range suggested by \citet{Fischera2003} in studying the starburst galaxy reddening and extinction based on a turbulent ISM model, and this is also the value adopted in jet-clouds simulations in \citet{Sutherland2007}, and \citet{Wagner2011, Wagner2012, Wagner2013}. The mean density $\mu$ is set to be $1.0 \times 10^3$ or $5.0\times 10^{2} ~\mhcm$ in our simulations (Table 1). 

The isotropic power spectrum D(k) is set to be a power-law dependence on k, to describe a self-similar structure:
\be
D(k)=\int{k^2 F(\textbf{k}) F^{\star}({\textbf k}) d \Omega} \propto k^{\beta}
\ee
where $F(\textbf k$) is Fourier transform function of the density field $\rho(\textbf r)$ with {\bf k} representing the wave vector, $F^{\star}({\textbf k})$ is the complex conjugate, and $d\Omega$ is solid angle element. We set $\beta=-5/3$, which represents the Kolmogorov spectrum. We set $k_{min}=5$ in a cubic box for generating clouds with side length of 250 meshes (corresponding to 50 pc), and therefore the maximum size of clouds is 250/(2$k_{min}$)=25 meshes (corresponding to 5 pc).
We use the python package -- pyFC\footnote{\url{https://pypi.python.org/pypi/pyFC}} to generate ``fractal cubes'', which is a fractal cube construction and analysis module written by Dr. A. Y. Wagner. 

 \begin{figure*}[!htb]
   \centering
   \begin{center}
   \includegraphics[width=0.4\textwidth]{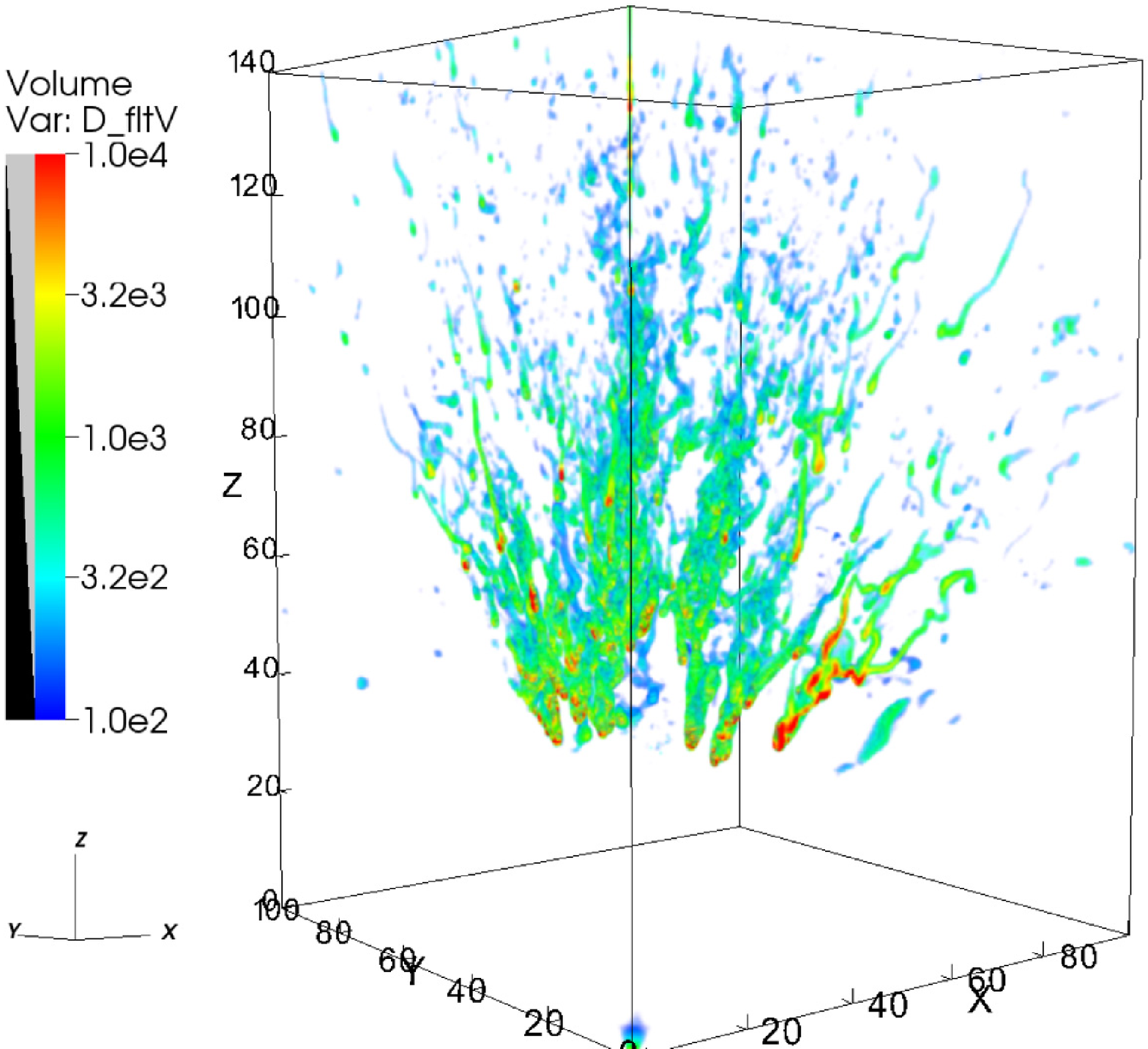}  \hspace{0.3cm}
   \includegraphics[width=0.4\textwidth]{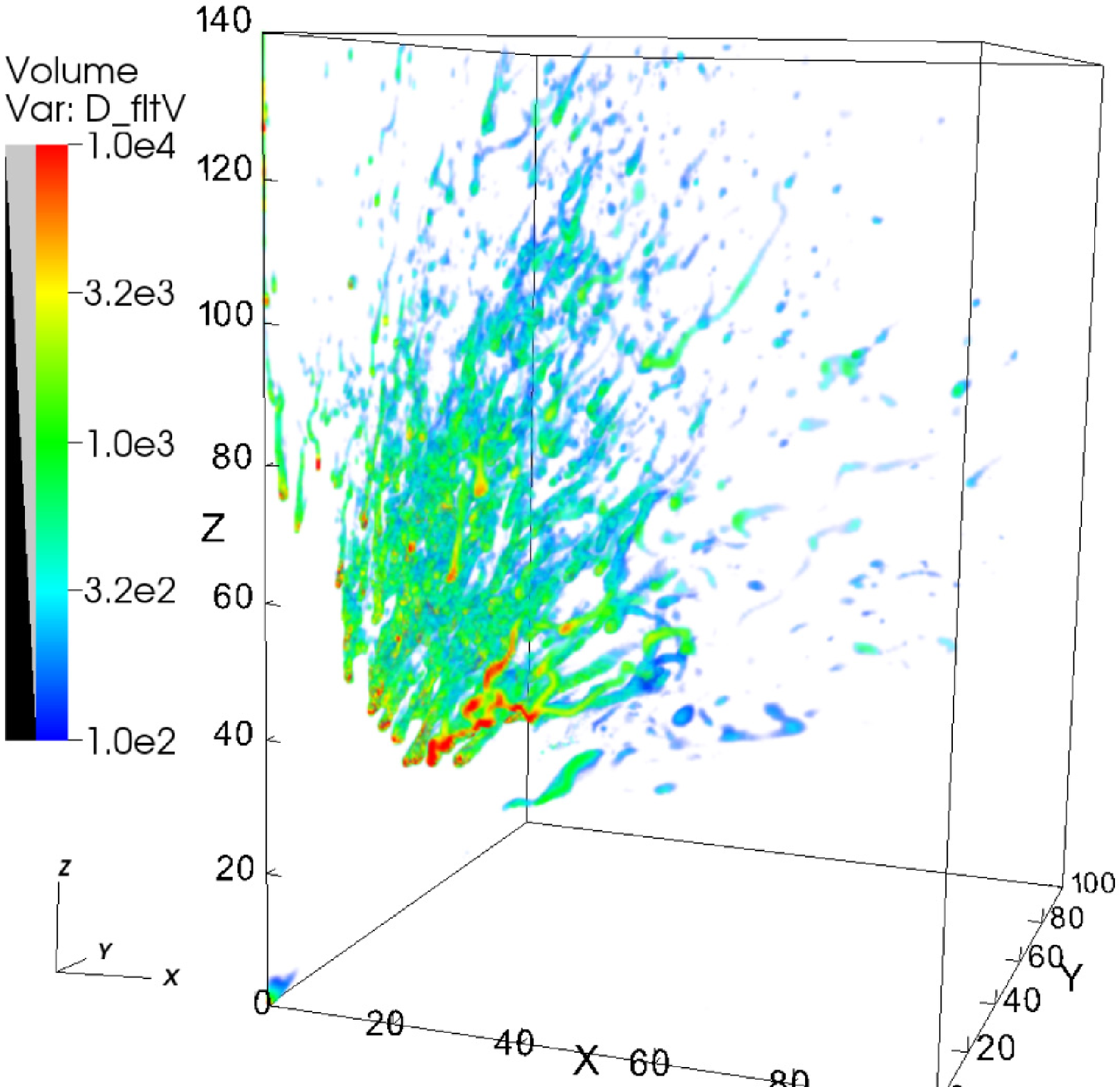} \vspace{0.5cm} \\ 
   \includegraphics[width=0.3\textwidth]{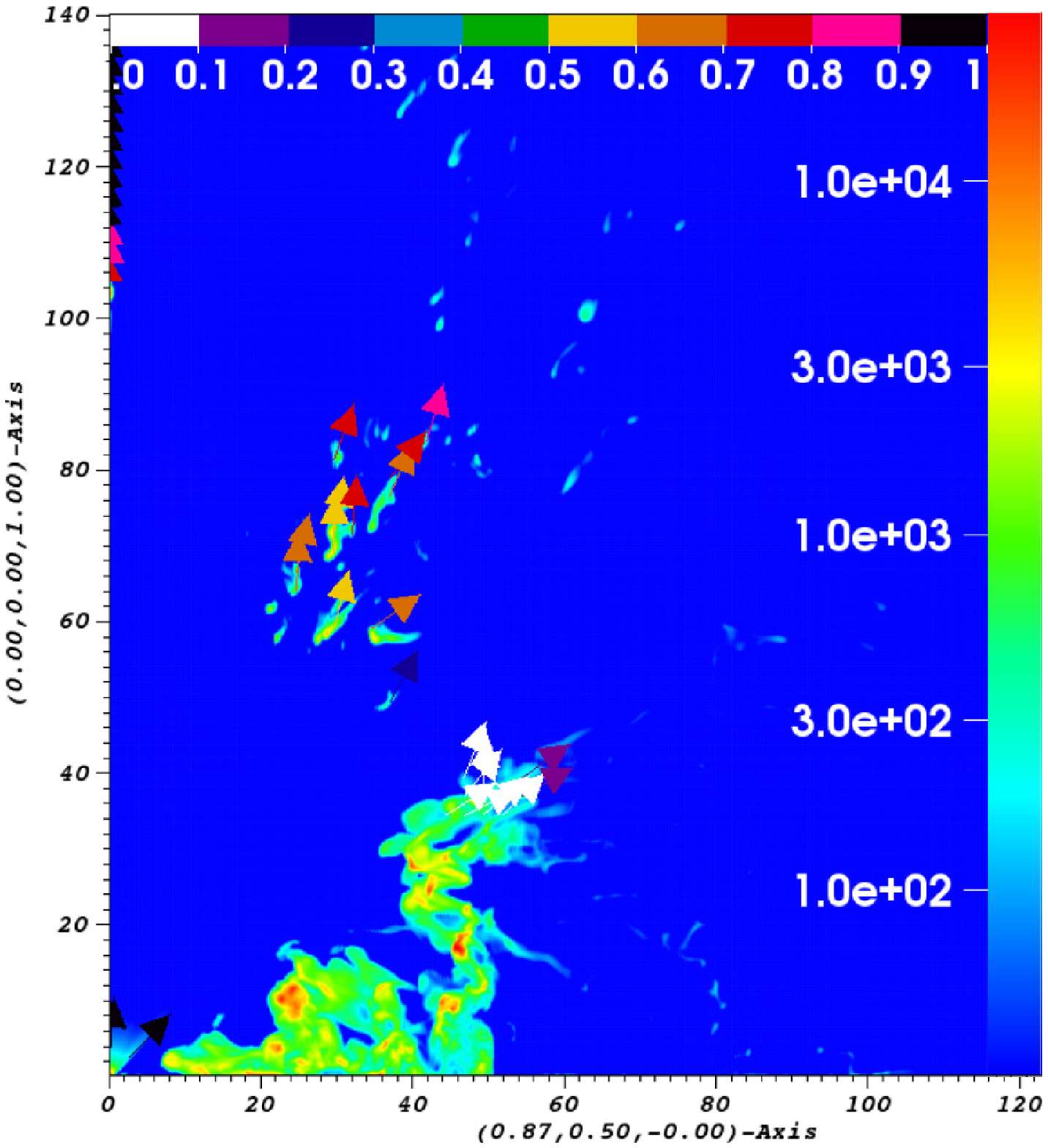} 
   \includegraphics[width=0.3\textwidth]{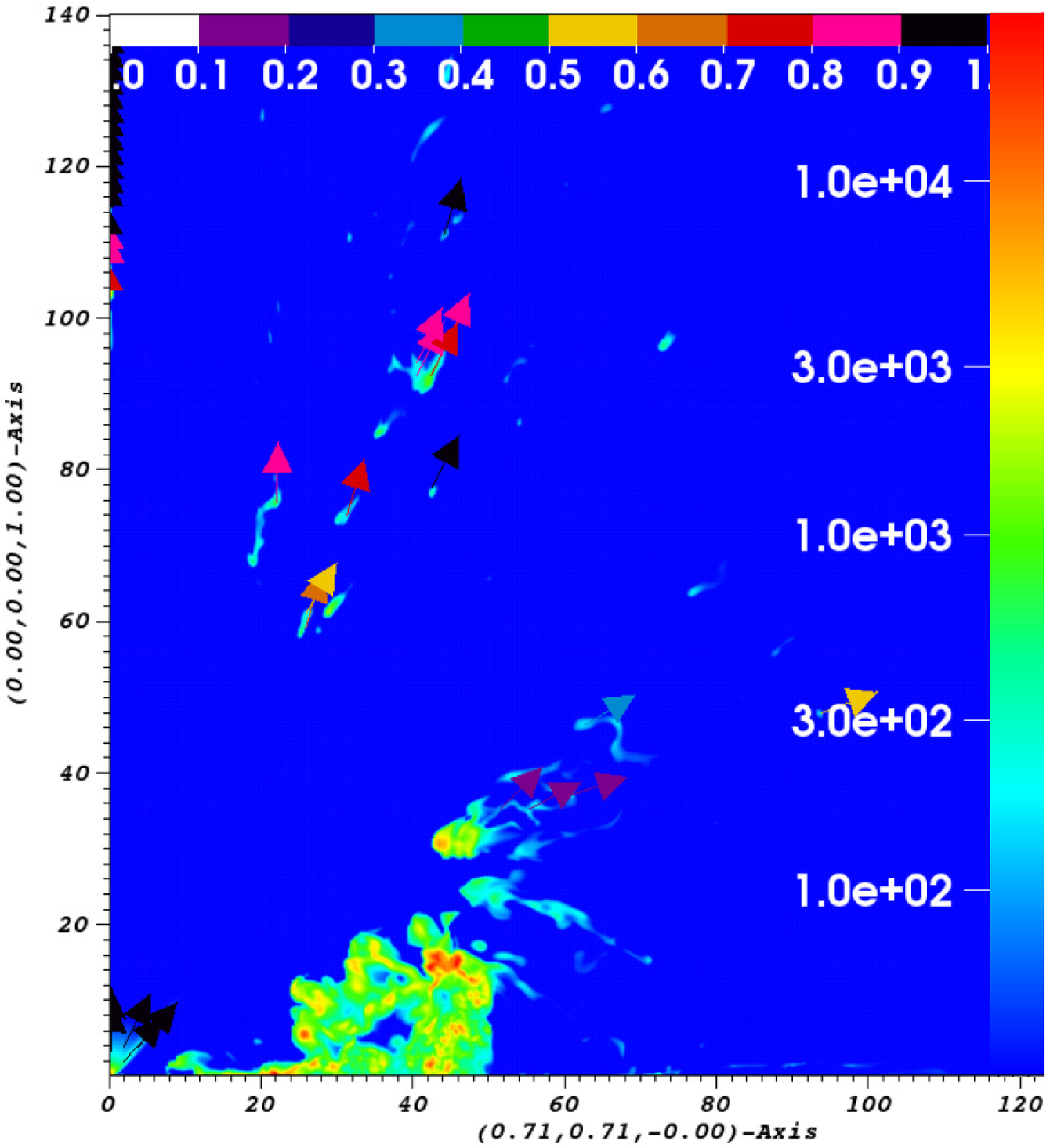}
   \includegraphics[width=0.3\textwidth]{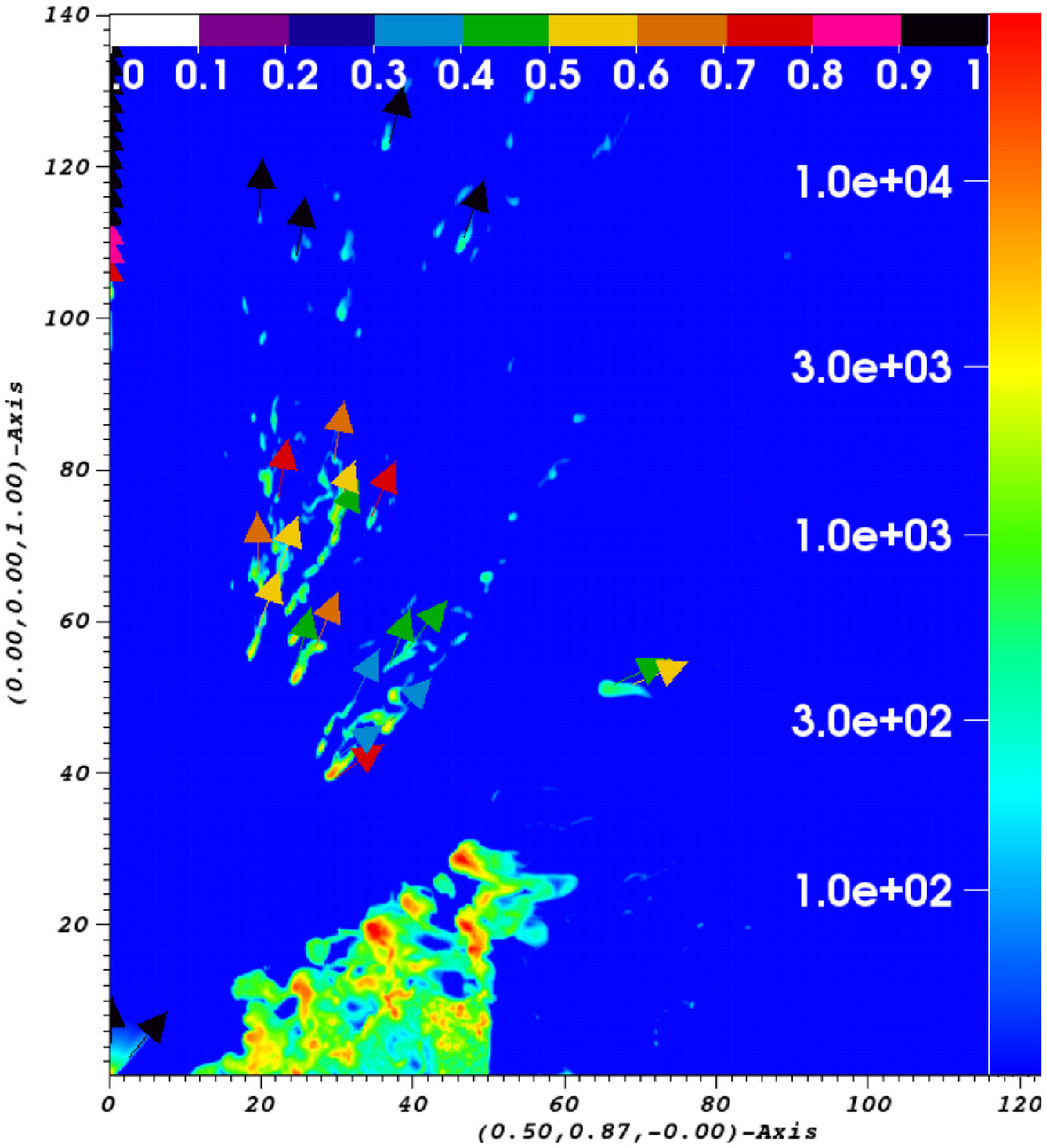}
  \end{center} 
   \caption{Results of run A. Top: 3D views of the density field filtered by $v_r \geq 100~\kms$ and $\rho \geq100~\mhcm$ at t=140 kyr, with different view angles. Bottom: density slices on the plane of $\phi=30^{\circ}$(\emph{left}), $45^{\circ}$(\emph{middle}), and $60^{\circ}$(\emph{right}) at 140 kyr. Vertical colorbar in each panel represents the density in units of $\mhcm$, and the horizontal discrete colorbar in lower three panels represents the velocity in units of $1000~\kms$.}
   \label{plot2}
  \end{figure*}

We impose an upper temperature cutoff $T_{cut}$ for the clouds, above which thermal instability arises. The temperature range for thermal instability is complex to calculate. We here simply set $T_{cut}$ to be $1\times10^4$ K, similar as $3\times 10^{4}$ K in \citet{Wagner2012}. We set a pressure equilibrium between the hot gas and clouds, and the lowest density in clouds is therefore $4 \times 10^{2} ~\mhcm$. 

As revealed by $H_{2} ~\lambda 2.1218\mu m$ radiation, circumnuclear clouds stretches to 1 arcsec in the direction of North-West and South-East direction, with a thickness of $\sim 1.5''$, and it stretches roughly perpendicularly to the cone axis of NLR (\citealt{StorchiBergmann2009, StorchiBergmann2010}). In our simulation domain, the extension of the region with circumnuclear clouds is set to be 50 pc and the height is 35 pc (see Figure 1). Next, in the circumnuclear clouds, we set a hollow cone centered at $Z$-axis with half opening angle of $20^{\circ}$ to mimic the morphology of torus (\citealt{Wada2009}), and such a cone-shaped hollow structure also seems to exist in the $H_{2} ~\lambda 2.1218\mu m$ image (Figure 6 in \citealt{StorchiBergmann2009}). This pre-existing cone may be generated in star formation process (\citealt{Wada2009}), or a past activity of the SMBH. 
The total mass of the clouds in eight octants is $1\times 10^7\msun$, which is roughly consistent with the total molecular gas mass of $\sim 10^8 \msun$ estimated in \citet{StorchiBergmann2010}.

In our simulation, we simply inject disk winds from the inner grids of $r \leq 1.5$ pc, and the density scales as $r^{-2}$ inside this injection zone. The densities of disk winds at $r=1$ pc are listed in Table 1. Disk winds are constrained within a cone with half opening angle of 40$^{\circ}$, and both the density and velocity do not vary with polar angles. Such cone-shape winds can be generated by the constraint on larger opening angle winds by the torus. 
For our basic run (run A), the mass outflow rate of disk winds is set to be 1.2 $\msunyr$, which is about 1.2 $\medd$, and the kinetic luminosity is $2.4\times 10^{43} ~\ergs$, corresponding to 0.3$\times$ bolometric luminosity. We also test two cases with lower kinetic luminosities, as shown in Table 1.


\section{RESULTS}
\subsection{Physics of NLR Outflows}

 \begin{figure*}
   \centering
   \includegraphics[width=0.85\textwidth]{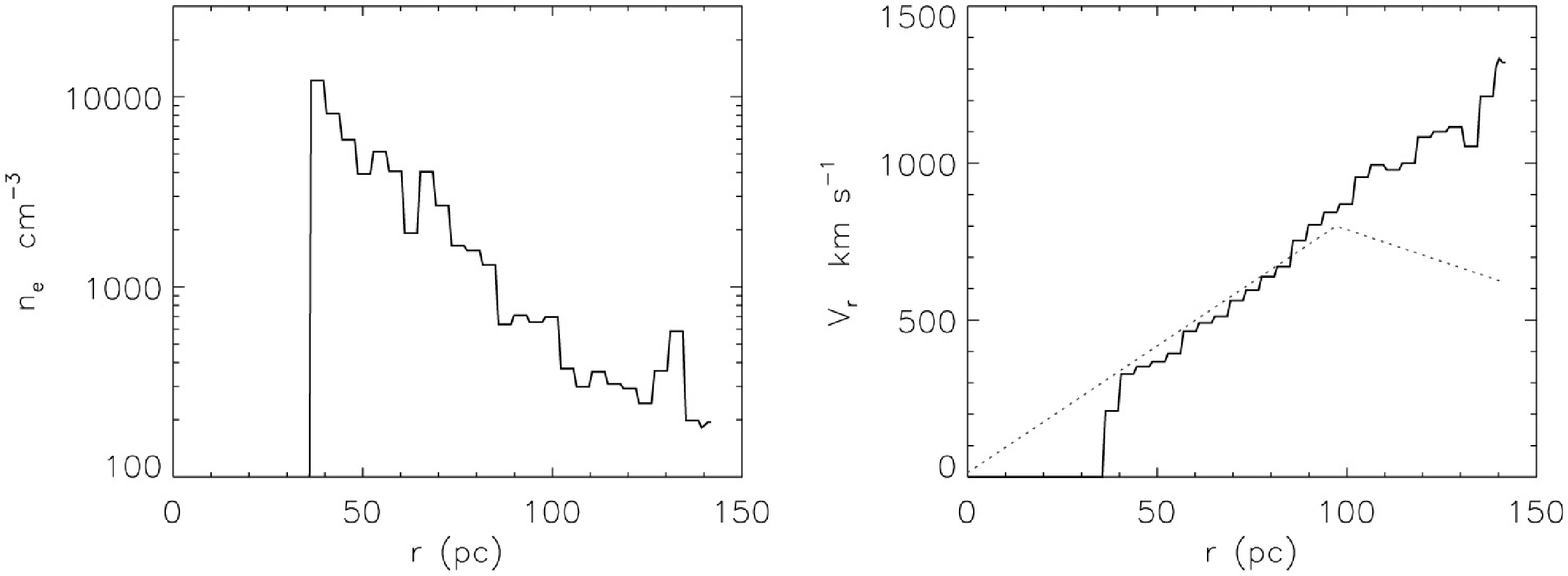} \hspace{0.0cm} \\
   \includegraphics[width=0.8\textwidth]{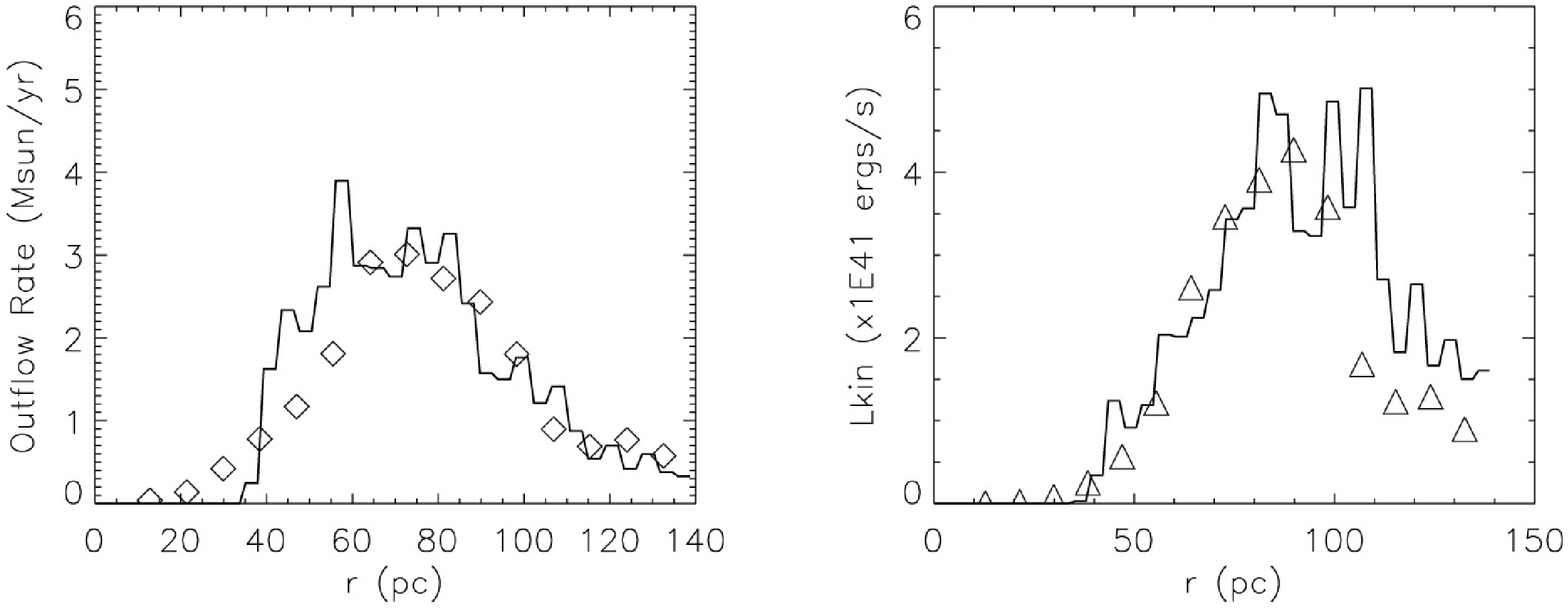}  \hspace{-0.1cm}
   \caption{Top: the \emph{left} panel shows the averaged density distribution of $n_e$ of run A at t=140 kyr for NLR outflows; the \emph{right} panel shows the averaged velocity with dotted line overplotted representing the result in \citet{Das2005}. Bottom: in the \emph{left} panel, the solid line shows the simulated mass outflow rate, and diamonds show the maximum confidence data of outflow rate (\citealt{Crenshaw2015}); in the \emph{right} panel the solid line shows the simulated kinetic luminosity, and triangles show the maximum confidence data of kinetic luminosity (\citealt{Crenshaw2015}). }
   \label{plot3}
  \end{figure*} 
 
We regard run A as our basic run, and mainly analyze the result of run A. 
In Figure \ref{plot2}, we plot 3D views of NLR outflows filtered by $\rho \geq100~\mhcm$ and  $v_r \geq 100~\kms$ at $t=140$ kyr, and show three slices of $\phi=30^{\circ}$, $45^{\circ}$, and $60^{\circ}$. Driven by disk winds, the circumnuclear clouds move outwards in a cone. Irradiated by the radiation from the accretion disk, these clouds can be photoionized and produce emission lines, like [O III]. 
We recognize the NLR outflows to be detected if the following conditions are satisfied:
(A) the clouds confined inside the ionization cone with $\theta \leq 33^{\circ}$ (\citealt{Das2005}); 
(B) density $\rho \geq 100~\mhcm$;
(C) distance larger than 10 pc.
Clouds outside the ionization cone will not be irradiated by the central source and therefore will not be photoionized to produce emission lines, and gas with density lower than $10^{2}~\mhcm$ will be difficult to detect since emissivity is proportional to $\rho ^2$. Combining condition B and C, disk winds themselves are excluded to be regarded as NLR outflows.

As shown in Figure \ref{plot2}, clouds with higher density move slower. For example, at $r \sim 50$ pc, clouds typically show density $\rho \sim 10^{4}~\mhcm$ and outflow velocity $v_{r} \sim 300-400~\kms$, while at $r\sim 100$ pc, clouds show a density of $\la 10^{3}~\mhcm$ and a velocity of $\ga 800~\kms$. Gas with higher density is more likely buried in larger clouds. Considering that acceleration is inversely proportional to density and the size of clouds, it is natural to comprehend the picture that clouds with higher densities move with lower velocities, resulting in their locations closer to the SMBH. 
However, it does not mean lower density clouds definitely can not appear in the innermost tens of parsecs. If clouds can continually move into the cone of disk winds in inner region (e.g., converging towards the $Z$-axis), which is not considered in our simulations, some clouds with lower densities and high velocities will emerge there. 

We plot the distribution of averaged density and averaged velocity in radial direction in Figure \ref{plot3}, and both of these parameters are averaged by weight of $\rho ^2$. These two parameters seem roughly consistent with the result of \citet{Kraemer2000}, \citet{StorchiBergmann2010} and \citet{Das2005}. 
We also note the anomalous data in $r\la 30$ pc where no NLR outflow exist. With high ram pressure in the inner part, the disk winds efficiently clear up the clouds by pushing them further away or destroying them, while there are no more clouds replenished. Considering a random motion of initial clouds may significantly improve the simulation results.

Besides, the simulated velocity significantly deviates from \citet{Das2005} at $r>100$ pc where they argued that velocity turns over to decrease with distance. If Das's trend is reliable (but see also \citealt{StorchiBergmann2010}), $\dot v_{r}$ for the gradual deceleration would be a few times higher than the gravitational acceleration, assuming $g \sim 2 \sigma^{2}/r$, in which $\sigma \sim 116 ~\kms$ for NGC 4151 (\citealt{Bentz2006}). Therefore additional forces pulling back or blocking the clouds should exist. \citet{Everett2007} and \citet{Das2007} argued that such a deceleration is mainly caused by resistance of ISM. However in our model, ISM is cleaned up at very early stage, and hence will not decelerate NLR outflows. 
We argue that deceleration may be caused by colliding with other clouds located at $r \ga 100$ pc. 
In the meanwhile, we also note that there is an UV absorber called component A in \citet{Kraemer2001}, with a blueshift of $\sim 1500 ~ \kms$, and density of $10^{2} ~\cmc$. The distance of this absorber is a few hundreds of parsecs from the center, which is just located in the scale of NLR (\citealt{Kraemer2001}). The velocity of component A seems to conflict with the Das's velocity distribution, but is consistent with our continuous acceleration feature. 
Therefore kinematics of NLR outflow beyond 100 pc is complex, and we remain this problem to future studies.

Shocks can destroy dust grains in clouds, and release refractory elements, like iron. Therefore, [Fe II]/Pa$\beta$ flux ratio can be regarded as an indicator of the dominant ionization mechanism -- photoionization or shock excitation (e.g., \citealt{StorchiBergmann1999}). 
We note that for NGC 4151, [Fe II]/Pa$\beta$ flux ratio from near-IR observations shows evidence for shock excitation at distance of $\ga 0.6$ arcsec (\citealt{StorchiBergmann2009}). In quality, this result is consistent with our model, since NLR outflows inevitably suffer from shocks by disk winds.

Finally, following \citet{Crenshaw2015}, we calculate the mass outflow rate $\dot M_{out}(r)$ and kinetic luminosity $L_{kin}(r)$ of NLR outflows at each radius by:
\begin{gather}
\dot M_{out}(r)=dM~\bar v_{r}/dr ,  \\  
L_{kin}(r)=\frac{1}{2} \dot M_{out}(r) \bar{v}^{2}_{r} ,
\end{gather}
where $dM$ is the mass inside the shell of $r\sim r+dr$, $\bar{v}_{r}$ is averaged by the weight of $\rho^{2}$ at distance of $r$. We checked another definition: $\dot M_{out}(r)=\oint{\rho v_{r} r^{2} \sin \theta d\theta d\phi}$, and found that the discrepancy is negligible.  
As shown in the lower panels in Figure \ref{plot3}, distributions of mass outflow rate, and kinetic luminosity in radial direction are roughly consistent with the results of \citet{Crenshaw2015}, except the inner part of $r\la 30$ pc. 

The kinetic luminosity of disk winds is almost two orders of magnitude higher than the observed NLR outflows. Therefore, only a small portion of the kinetic energy of disk winds can be transferred into the clouds. The vast majority is carried by hot and dilute winds flooding through no-cloud or inter-cloud channels, which shows a density of $\la 10^{0}~\mhcm$, much lower than clouds. The shocked disk winds in NLR region are undetectable, because of the low density and the high temperature (which is as high as $10^{8-10}$ K in post shock region). The existence and strength of disk winds can be confirmed by their impacts on the ISM with high densities, e.g., shock excitation signs as mentioned above, and the pressure in NLR as discussed in section 5 below. 

There are two mechanisms for driving ISM -- momentum driven and energy driven, and the key distinction between them is whether cooling of shocked winds is enough efficient or not (\citealt{Faucher-Giguere2012}). For driving clouds as in our case, adiabatic cooling dominates the cooling of shocked winds. This cooling is significant, but not enough efficient, especially at the head of clouds which is the main action area for accelerating clouds. 
And we also find that at early stage, the momentum of clouds indeed exceeds that of injected disk winds (e.g., at 40 kyr for run A, the momentum is $9.4\times 10^{46}~{\rm g~cm~s^{-1}}$ for the clouds while it is $7.9\times 10^{46}~{\rm g~cm~s^{-1}}$ for the total disk winds). Therefore we argue that clouds are between momentum and energy driven, which is consistent with the conclusion in previous jet-clouds simulations (\citealt{Wagner2012}).

  \begin{figure*}[!htb]
   \centering
   \includegraphics[width=0.48\textwidth]{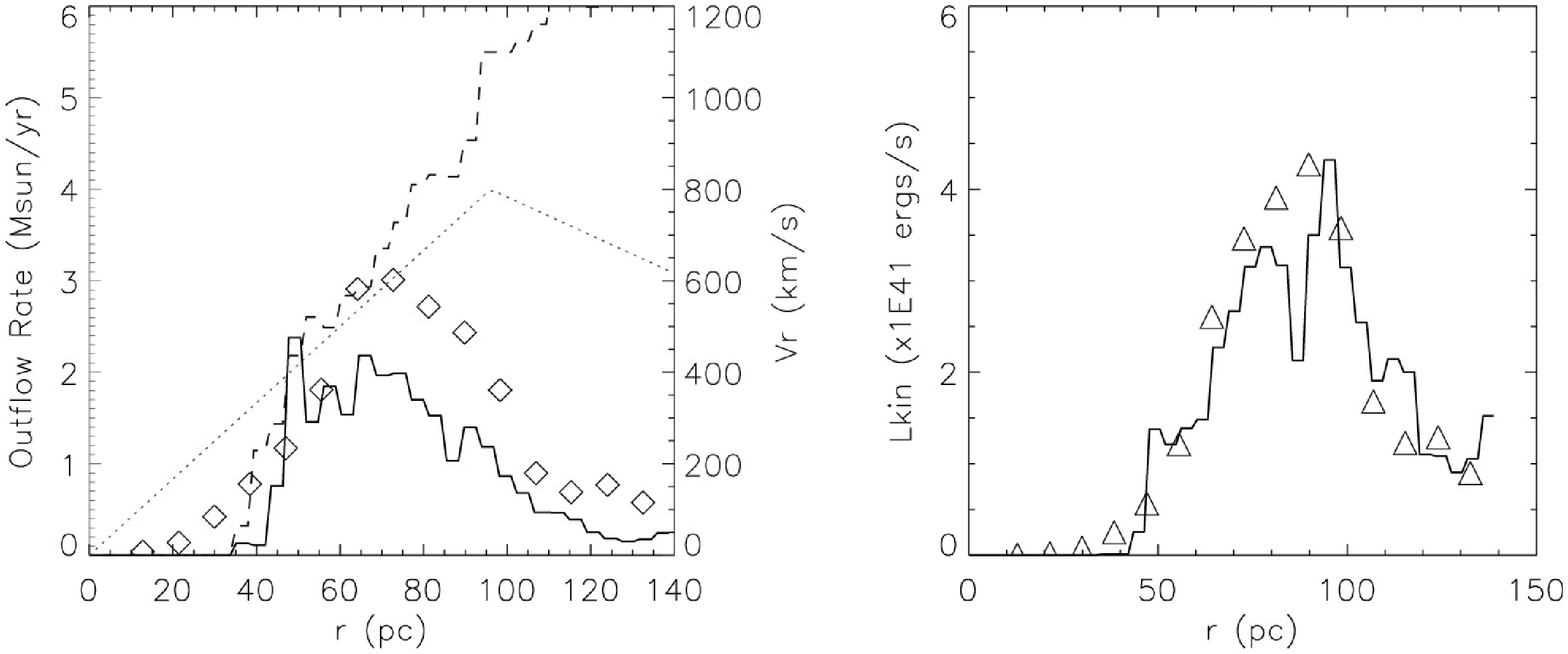} \hspace{0.5cm}
   \includegraphics[width=0.48\textwidth]{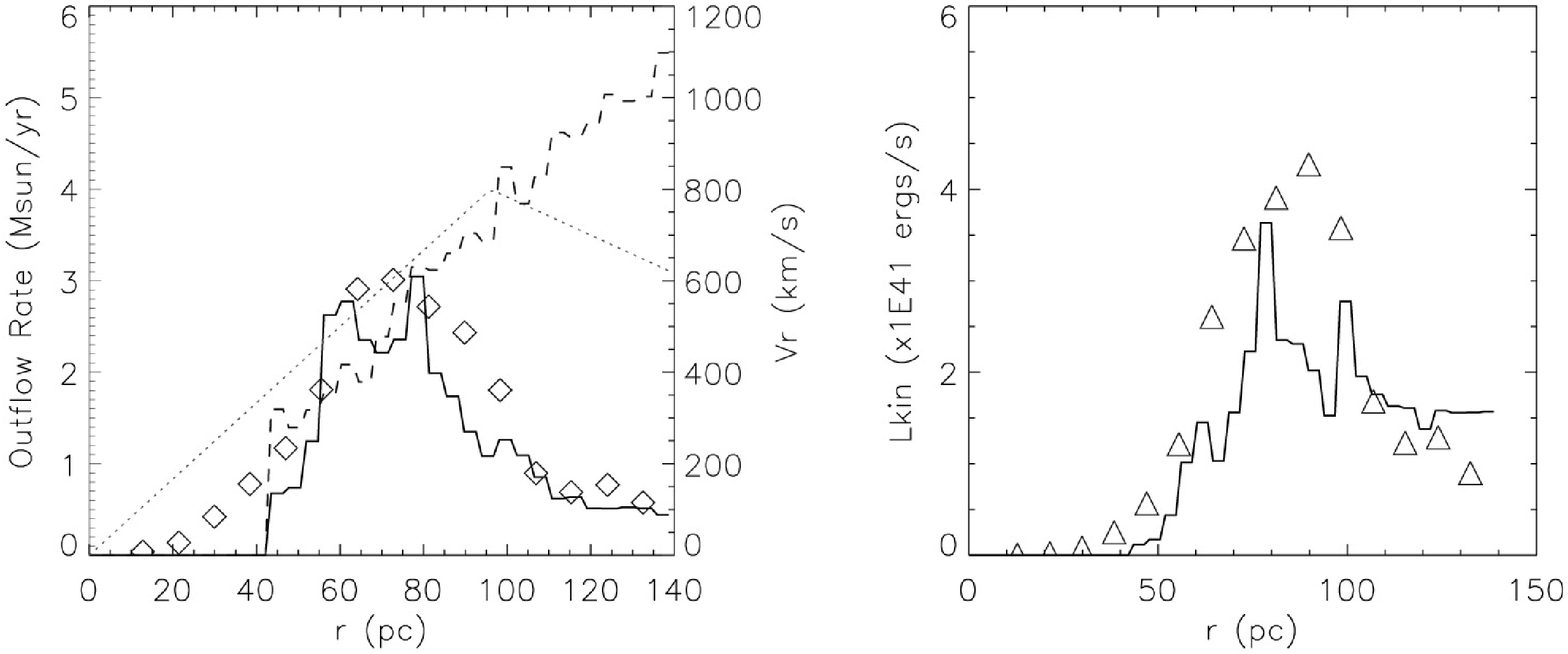}  \\
   \includegraphics[width=0.48\textwidth]{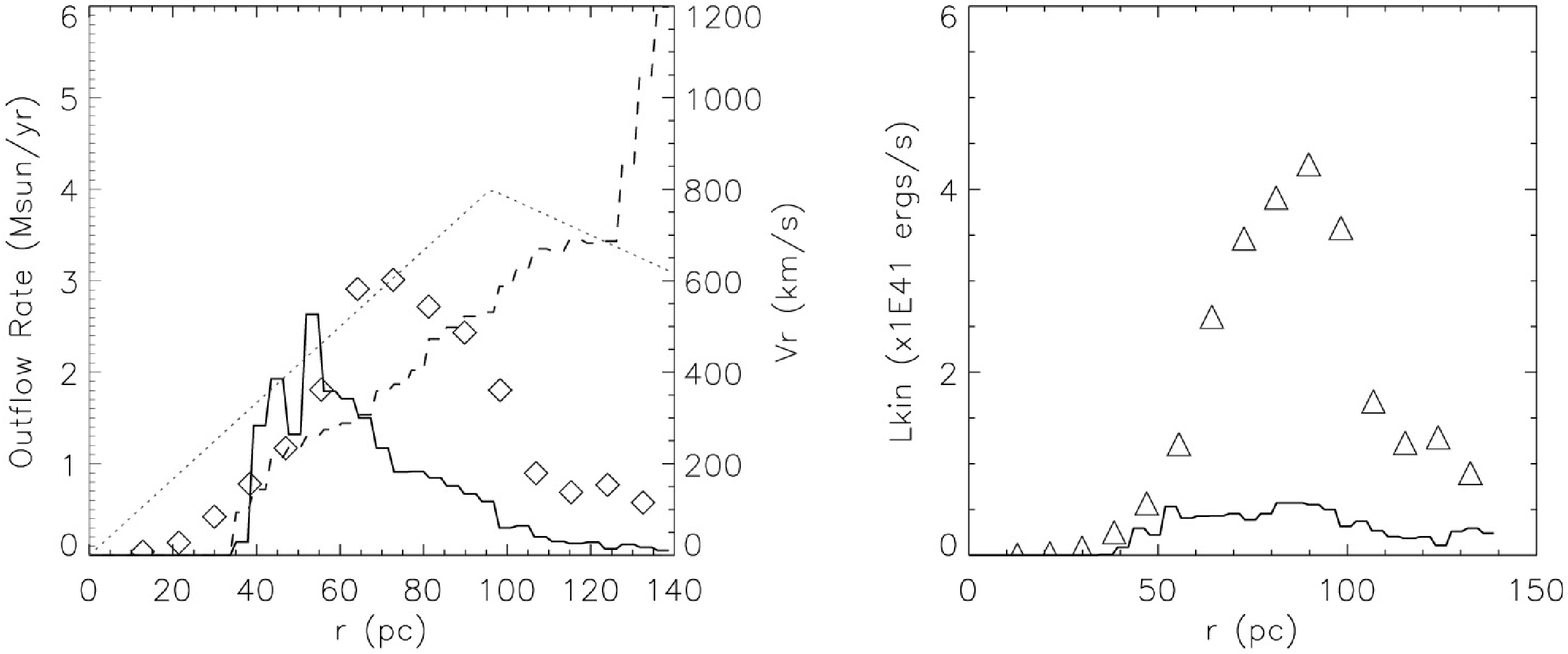} \hspace{0.5cm}
   \includegraphics[width=0.48\textwidth]{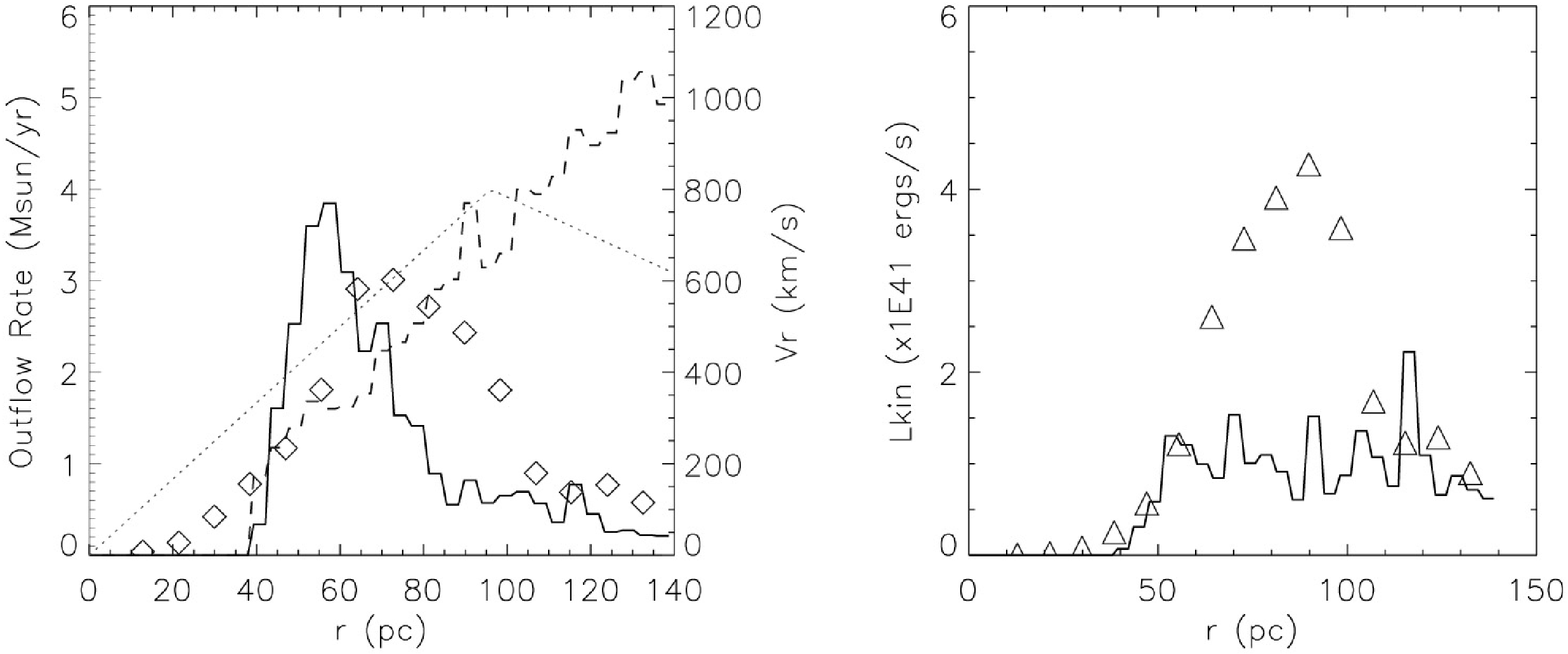}
  \caption{Distributions of mass outflow rate (solid lines), velocity (dashed lines) and kinetic luminosity (solid lines) in radial direction: \emph{top left} two panels -- run B, \emph{top right} two panels -- run C,  \emph{bottom lefts} -- run D,  \emph{bottom rights} -- run E. We use diamonds and triangles to show the maximum confidence data of outflow rate and kinetic luminosity respectively as suggested in \citet{Crenshaw2015}, and plot dotted line to represent the velocity of \citet{Das2005}. }
   \vspace{0.3cm}
   \label{plot4}
  \end{figure*}


\subsection{The Effects of Changing Parameters}
We also test the effects of changing parameters, and plot the distributions of mass outflow rate, velocity and kinetic luminosity of NLR outflows in Figure \ref{plot4}.

In run B, the mean density of the clouds is reduced by one half. We find that, the mass outflow rate of NLR outflows is significantly lower than run A, while the velocity is slightly higher. Acceleration is inversely proportional to the cloud density, and thus clouds inside the injection cone of disk winds can be driven outwards more efficiently in this lower density case. 

In run C, we test a case of slow disk winds, with a velocity of $4000~\kms$. The ram pressure $\rho v^2$ of disk winds is kept unchanged, compared with run A. We find that, evolution time is longer than run A to obtain a similar result. This implies that, driving clouds by disk winds is not a pure momentum-driven mechanism, since the momentum rate is $\dot P_{\rm wind} \propto \rho_{\rm wind} v^2_{\rm wind}$, and is same as run A. Otherwise, the evolution time should be the same. This agrees with the discussions in section 4.1.

In run D, we test a case of slow and weak disk winds, in which we keep the velocity of $4000~\kms$, but reduce the density of disk winds by 75\% compared with run C. It is obvious that, the velocity of NLR outflows is significantly lower than run C.

In run E, we replace the initial circumnuclear clouds with larger clouds, of which the maximum size is 12.5 pc (it is 5 pc in the other runs). Since the acceleration is inversely proportional to the cloud size, the velocity is significantly lower in this case, compared with run A. 

\section{DISCUSSION}
 \begin{figure}
   \includegraphics[width=0.45\textwidth]{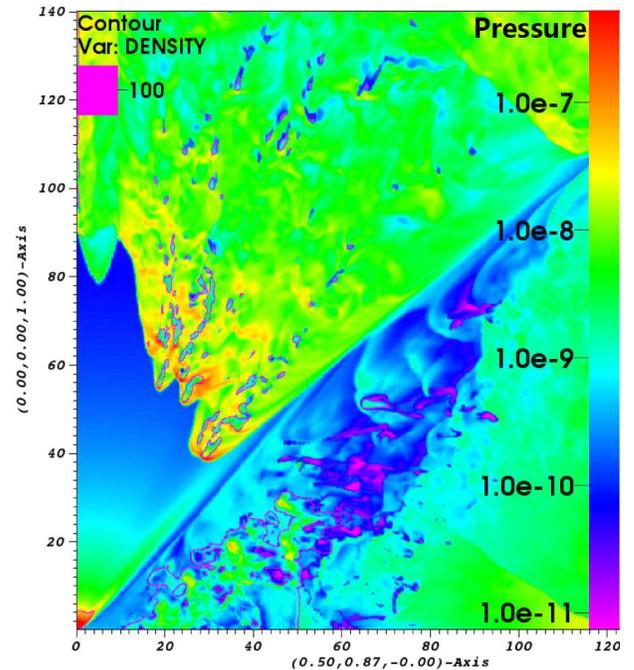}
   \caption{Thermal pressure distribution of run A in slice of $\phi=60^{\circ}$ ($t=140$ kyr). The pressure is in units of dyne cm$^{-2}$. The thin magenta contours show the density of 100 $\mhcm$. }
   \label{plot5}
  \end{figure} 

The profile of thermal pressure is shown in Figure \ref{plot5}. The most conspicuous feature is that thermal pressure is highly inhomogeneous, with $P_{th}= 10^{-8}-10^{-9}~\dyncm$ inside the NLR clouds, $10^{-7} ~\dyncm$ in shocked winds ahead of NLR clouds, and $10^{-7}-10^{-10}~\dyncm$ in the free expanding supersonic disk winds. 
According to spectral fitting of optical/UV band emission for NLR by photoionization model, the density and temperature of NLR (photoionized clouds) is $1000~\cmc$, and $1.3-1.9 \times 10^{4}$ K, respectively (\citealt{Alexander1999}). Therefore, thermal pressure inside these NLR clouds is $\sim 10^{-9} ~\dyncm$, which is further confirmed in the study of fitting the soft X-ray spectrum of X-ray enhanced region by a collisional ionization model, in which the temperature is 0.58 keV and thermal pressure is 6.8$\times 10^{-10} ~\dyncm$ (\citealt{WangJF2011b}).
Besides, from several emission lines observed at the locations where the [Fe II] emission is enhanced, \citet{StorchiBergmann2009} have obtained the gas density in NLR, $n_{e} \approx 4000~\cmc$, and temperature $T_{e} \approx 15 000 \pm 5000$ K, which means a total thermal pressure of $\sim (1.7 \pm 0.6) \times 10^{-8}~\dyncm$ assuming $P_{\rm electron} \approx P_{\rm ion}$. Therefore, observations show that thermal pressure inside the NLR clouds is in the range of $10^{-9}-10^{-8}~\dyncm$, which is close to our simulation result. Besides, there are radio knots (usually regarded as jet structures) located at $1''$-$2''$ away from the nucleus in 5-GHz and 8-GHz map (marked as C1, C2, C5 in \citealt{Pedlar1993}), which are close to the X-ray enhanced region (\citealt{WangJF2011c}). The total pressure in these radio knots, which is dominated by high energy electrons and magnetic field, is $10^{-7}~\dyncm$ according to synchrotron radiation model (\citealt{Pedlar1993}). 
This pressure is consistent with the thermal pressure of shocked winds ahead of NLR clouds and the ram pressure of disk winds at $\sim 100$ pc in our model.

The total pressure at a distance of $\sim$ 100 pc suggested in radio maps is comparable to the ram pressure of disk winds in our model, which actually characterizes the strength of disk winds. It is impossible for a weak wind of $10^{41}~\ergs$ to create such a high pressure, unless the velocity of the weak wind is as low as $\sim 100~\kms$. This implies that the observed NLR outflows may be not the whole story, and another underlying component probably coexist, which is the disk winds or shocked disk winds in our model.

The present radiation is not important in driving the circumnuclear clouds. The momentum rate of photons $L_{bol}/c$ is several times lower than the NLR outflows, and therefore is not enough for driving NLR outflows even if the circumnuclear clouds are optically thick. \citet{WangJF2010} have studied the formation of extending soft X-ray emission filling in the $\sim 2$ kpc scale HI cavity around the nucleus. To explain the strong X-ray emission with photoionization model, they suggested that a quasar state of which the ionizing luminosity is close to the Eddington luminosity should have occurred several $10^{4}$ years ago (shorter than our simulation times), while subsequently it weakened. If so, radiation driven mechanism for quasar state happened in the past may play an important role in forming NLR outflows, which remains to study in the following work. 

Although the kinetic luminosity of disk winds is roughly in the same order of magnitude as that of UFO, it is as high as 30\% of the present bolometric luminosity, which seems too high (e.g., \citealt{Gofford2015}; \citealt{Nomura2017}). This may be the most controversial point of our disk wind driven model.
We argue that, if NGC 4151 did experience a quasar state $10^4$ years ago (\citealt{WangJF2010}), the ratio of the kinetic luminosity of disk winds to the bolometric luminosity would be much lower, which should be in the order of a few 0.1\%. On the other hand, when some physics is considered, the required kinetic luminosity of disk winds can be reduced. 
When magnetic field is included, drag force on clouds would be enhanced by $1+(v_{\rm A}/v_{\rm wind})^2$ compared with the hydrodynamic force where $v_{\rm A}$ is the Alfven speed defined as $v_{\rm A} \equiv B/\sqrt{4\pi \rho}$ (\citealt{Dursi2008}; \citealt{McCourt2015}). 
Recent MHD simulations also prove that for clouds driven by jet/winds, the magnetic tension force from the stretched field lines can indeed promote the acceleration of these clouds (\citealt{Asahina2017}). 
When viscosity is considered, high velocity disk winds can exert viscous torque on the clouds, which acts as another driven force (relative to the ram pressure) to accelerate the clouds. 
Besides, motion of clouds in initial conditions is not included in our simulations. The tails of clouds dragged by winds may suffer a stronger ram pressure when motion of clouds is considered. Taking into account the above factors, the kinetic luminosity of the disk winds suggested in our basic simulation should be regarded as an upper limit.

\section{SUMMARY}

Selecting NGC 4151 for study, we have performed 3D hydrodynamic simulations to study the formation of NLR outflows. We assume that, these NLR outflows are formed by circumnuclear clouds driven by disk winds, and explore the parameters of such underlying disk winds. We summarize our main results as below.

\begin{itemize}

\item Our model can explain the trend of outflow velocities in inner 100 pc, but can not explain the deceleration trend outside 100 pc yet. 

\item In order to explain the properties of the NLR outflow, strong disk winds are required. The velocity of such disk winds is a few thousands $\kms$, and the kinetic luminosity of disk winds is two orders of magnitude higher than the kinetic luminosities of NLR outflows in $10^{1-2}$ pc scale, but comparable to that of UFO, and a few times lower than the present bolometric luminosity. 

\item 
Only a small portion of accretion wind energy can be transferred into the clouds to form NLR outflows, while the vast majority is carried by hot and dilute winds flooding through no-cloud or inter-cloud channels. The density of these dilute winds in NLR is $\la 10^{0} ~\mhcm$, and the temperature of post shock winds is as high as $10^{8-10}$ K. Therefore disk winds in NLR scale should be too dilute or too highly ionized to generate emission and absorption lines. 
In inner parsec or sub-parsec region, nondetection of disk winds may be due to the high ionization parameter, or obscuration by the putative torus or the dusty inner galactic disk.
The existence of the underlying disk winds can be confirmed by their impacts on clouds, e.g., shock excitation signs, and the pressures in NLR. 

\item Thermal pressure of the NLR clouds in simulations is consistent with observations. Thermal pressure in shocked disk winds is also consistent with the total pressure inferred from radio knots. 

\end{itemize}

\acknowledgements
We thank A. Y. Wagner for the help in using pyFC to set initial clouds, and Feng Yuan for very helpful discussions. This work is supported by the Fundamental Research Funds for the Central Universities and the National Natural Science Foundation of China (NSFC--11421303). T.G.W. acknowledges support from National Basic Research Program of China (grant No.2015CB857005), NSFC (NSFC-11233002, NSFC-11421303). Simulations were made using the High Performance Computing Resource in the Core Facility for Advanced Research Computing at Shanghai Astronomy Observatory, and the supercomputing system in the Supercomputing Center of University of Science and Technology of China.

\end{document}